\documentclass[11pt]{article}

\usepackage[a4paper, textwidth={17cm}]{geometry}

\usepackage{amsfonts}
\usepackage{amsmath}
\usepackage{amsthm}
\usepackage{amssymb}
\usepackage{authblk}

\usepackage{algorithm}
\usepackage{algorithmic}

\usepackage{graphicx}

\usepackage{color}
\usepackage[usenames,dvipsnames,svgnames,table]{xcolor}

\usepackage{bbm}

\usepackage[OT4]{fontenc}
\usepackage[utf8x]{inputenc}

\newtheorem{remark}{Remark}

\newcommand{\mket}[1]{| #1 \rangle}
\newcommand{\mbra}[1]{\langle #1 |}

\newcommand{\nC}{\mathcal{C}}
\newcommand{\nH}{\mathcal{H}}

\begin{document}

\title{Recommendation systems with quantum k-NN \\ and Grover's algorithms for data processing}

\author[1]{Marek Sawerwain}
\author[1]{Marek Wr\'oblewski}

\affil[1]{Institute of Control \& Computation Engineering 
	
	University of Zielona G\'ora Licealna 9, Zielona G\'ora 65-417, Poland
	
	e-mail: {M.Sawerwain@issi.uz.zgora.pl}, {M.Wroblewski@issi.uz.zgora.pl}}

\date{}

\maketitle
	
\begin{abstract}
In this article, we discuss the implementation of a quantum recommendation system that uses a quantum variant of the k-nearest neighbours algorithm and the Grover algorithm to search for a specific element in unstructured database. In addition to the presentation of the recommendation system as an algorithm, the article also shows a main steps in construction of a suitable quantum circuit for realisation of a given recommendation system. The computational complexity of individual calculation steps during recommendation system was also indicated. The verification correctness of a proposed recommendation system was also analysed, indicating an algebraic equation describing the probability of success of the recommendation. The article also shows numerical examples presenting the behaviour of the recommendation system for two selected cases.
\end{abstract}


\section{Introduction} \label{lbl:sec:Introduction}

The development of quantum computational methods \cite{Nielsen:2010} allows for its use in areas such as machine learning \cite{Wiebe:2015}, decisions models \cite{Busemeyer:2012} or big-data \cite{Nielsen:2016}. The classical methods of analysing a data sets of big-data are widely used, but the use of quantum computers that allow the processing of exponential amounts of data seems to be extremely important in this area \cite{Stefanowski:2017}, \cite{Veloso:2015}. In addition, due to the features of quantum information, the lack of the possibility of cloning information and direct comparison of two quantum registers, the construction of such a system requires a slightly different approach. In this article we show that the recommendation process can be based on the method of k-nearest neighbours classification \cite{Schliep:2004}. Such approach allows to create a system where the effectiveness of the recommendation can be very high.

In the article, we discuss the construction of a recommendation system based on two quantum computing algorithms. The first is the quantum algorithm of k-nearest neighbours (termed as qk-NN) \cite{Pinkse:2013}, \cite{Schuld:2014}, \cite{Wiebe:2015}, based on the method presented in the work \cite{Trugenberger:2002}. In this algorithm, we use the Hamming distance to classify the elements to be recommended \cite{Wisniewska:2018}. Grover's algorithm \cite{Grover:1996}, \cite{Erdal:2005} is used to improve the quality of recommendations. The use of both methods allows for the construction of a recommendation algorithm, which in the recommendation process is characterized by the better computational complexity than classical approaches \cite{Alpaydin:2004}, \cite{Armbrust:2010}. However, it should be emphasized that the described solution, like other quantum algorithms \cite{Shor:1999}, is probabilistic, however the probability of indicating recommended elements is very high.

It should be noted that this article is an extension of a publication \cite{Sawerwain:2019}. This article presents more precisely the structure of the quantum register and two numerical examples showing the behaviour of the recommending system. The computational complexity of individual steps implemented in the recommendation process was also indicated.

The organization of the article is as follows: in the part of \ref{lbl:sec:QI:Basics:AMCS2018:MS:MW} selected issues of quantum computing, such as quantum register and superposition are presented. The mentioned concepts characteristic of quantum computing are directly applied in the discussed recommendation system. In the part \ref{lbl:sec:QRS:AMCS2018:MS:MW} we discuss the construction of the recommendation system, present the algorithm of conduct and discuss the construction of the quantum circuit, which implements the discussed solution. The correctness of the recommendation system operation, i.e. the description of the probability of indicating the correct element is presented in the section \ref{lbl:sec:algebraic:correctness:AMCS2018:MS:MW}.
Numerical experiments were also implemented, which are discussed in the section~\ref{lbl:sec:numerical:experiment:AMCS2018:MS:MW}. The experiment shows the behaviour of the system in two cases when one and two elements are recommended. The summary of the article is given in the section~\ref{lbl:sec:conclusions:AMCS2018:MS:MW}, where the final conclusions and some comments about possible further work on the method described in this article were given. The article ends with thanks and a list of the cited bibliography.

\section{Brief introduction to the Quantum Information Basics} \label{lbl:sec:QI:Basics:AMCS2018:MS:MW}

The concept of a classic bit, which is the basic unit of information, can be extended to qubit for quantum systems \cite{Walther:2005}, \cite{Steane:1998}. For this purpose, we consider a two-dimensional Hilbert space $\nH_2$ and indicate orthonormal basis:

\begin{equation}
\vert 0 \rangle =
\left[
\begin{array}{c}
1\\
0\\
\end{array}
\right],
\enspace
\vert 1 \rangle =
\left[
\begin{array}{c}
0\\
1\\
\end{array}
\right].
\label{lbl:eq:std:base:AMCS18:MS:MW}
\end{equation}

In last equation a Dirac's notation has been used. The vector $\mket{0}$ usually means a zero. The natural thing is that $\mket{1}$ represents an one. The given Eq.~\ref{lbl:eq:std:base:AMCS18:MS:MW} base is also called a standard computation base.

In quantum information, notion of qubit is the equivalent of the classical bit. The qubit state is represented by a vector in the two-dimensional Hilbert space $\nH$:

\begin{equation}
\mket{\psi} = \alpha \mket{0} + \beta \mket{1}
\label{lbl:eq:qubit:unk:state:AMCS18:MS:MW}
\end{equation}

This vector is normalized, therefore $|\alpha|^2 + |\beta|^2 = 1$ occurs, wherein $\alpha, \beta \in \nC$ (where $\nC$ represents a set of complex numbers). The presented state $\mket{\psi}$ is called a vector state or a pure state.

Following the classic computer science, concatenation of multiple classical bits creates a register. The $\mket{\psi}$ register with $n$ -- qubits is built by using a tensor product:
\begin{eqnarray}
\mket{\psi} = \mket{\psi_0} \otimes \mket{\psi_1} \otimes \mket{\psi_2} \otimes ... \otimes \mket{\psi_{n-1}} , 
\label{lbl:eqn:product:state:AMCS18:MS:MW}
\end{eqnarray}
where $\otimes$ is a tensor product operation.

\begin{remark}{Quantum entanglement}
It should be added that there are cases where the register can not be written in the form of a tensor product. This state of the register is called a entangled state. $\square$
\end{remark}

It should also be noted that in addition to the description in the form of pure states, so-called unknown density matrices - the pure state of qubit $\mket{\psi}$ taking the following form are also used:
\begin{equation}
\rho = \mket{\psi}\mbra{\psi} = \left[ \begin{array}{cc} \alpha^2 & \alpha\beta \\ \alpha\beta & \beta^2 \end{array} \right]
\end{equation}
where the $\mbra{\psi}$ vector represents a transposed $\mket{\psi}$ vector.

In general, the representation for a mixed state for which only pure states are included: $\mket{\psi_i}$ takes the following form:
\begin{equation}
\rho = \sum_{i} \lambda_i \mket{\psi_i}\mbra{\psi_i},
\end{equation}
where $\lambda_i$ determines the probability of a state $\mket{\psi_i}$ and $\sum_i \lambda_i = 1$.

\begin{remark}{Exponential capacity of a quantum register}
One of the main differences between the classic and quantum registers is the exponential capacity of the quantum register. The amount of classical information contained in the quantum register described by the state vector can be described by $2^n$, where $n$ is the number of qubits. What causes that classical simulation of the quantum register is unfortunately not possible effectively using a classical computations methods. There are, of course, special cases such as the so-called CHP circuits \cite{Aaronson2004} states, but in general the quantum register requires an exponential computational resources to simulate the operation of the quantum register using a classical machine. This situation is much worse when we using a density matrix, as their size is described as $d^n \times d^n$.
	
The state vector for 16 qubits needs 512kb of memory, and the 256MB for density matrix, assuming that we use double precision floating point numbers describing individual amplitudes of probability. However, doubling the number of qubits to 32 due to the exponential capacity of the quantum register already requires 32GB of available memory, and 1 ZB (zetta bytes) for the density matrix. $\square$
\end{remark}

A several additional operations can be performed on the quantum register. As part of this brief introduction, we will only quote the two most important examples of operations, i.e. unitary operation and measurement operation.

The realization of unitary operation of $U$ for the quantum state represented by the vector is represented by the equation:
\begin{equation}
U \mket{\psi_{0}} = \mket{\psi_{1}}.
\end{equation}
For a density matrix, the unitary operation $U$ is described by the following relationship:
\begin{equation}
U \rho_0 U^{\dagger} = \rho_1 .
\label{lbl:eqn:density:matrix:unitary:operation}
\end{equation}
A very important thing is the method of creating a unitary operation for the quantum register. In the case of state modifications for the first and third qubit, the unitary unit construction takes the following form:
\begin{equation}
U = u_1 \otimes I \otimes u_2 
\label{lbl:eqn:tensor:product:of:u:gates:ms}
\end{equation}
The unitary operation is a reversible operation (called also as uncompute operation), i.e. for the operation $U$ you can always enter the operation $U^{\dagger}$, where the symbol $\dagger$ represents the Hermitian adjoint operation. If $U=U^{\dagger}$ then $U$ is called as a self-adjoint operation.

The basic set of unit operations includes a so-called Pauli's gates (operators): $X$, $Y$, $Z$. Their matrix representation is as follows:
\begin{equation}
X=\left(
\begin{array}{cc}
0 & 1 \\
1 & 0 \\
\end{array}
\right), \;
Y=\left(
\begin{array}{cc}
0 & -i \\
i & 0 \\
\end{array}
\right), \;
Z=\left(
\begin{array}{cc}
1 & 0 \\
0 & -1 \\
\end{array}
\right).
\end{equation}

The Hadamard gate marked as $H$ is also very important:
\begin{equation}
H = \frac{1}{\sqrt{2}} \left(
\begin{array}{cc}
1 &  1 \\
1 & -1 \\
\end{array}\right),
\end{equation}
because it is used to introduce the so-called superposition for quantum states. For example, let $\vert \Xi \rangle$ be $n$ - qubit state:
\begin{equation}
\vert \Xi \rangle = \vert \xi_{0} \rangle \otimes \vert \xi_{1} \rangle \otimes \dots \otimes \vert \xi_{n-2} \rangle \otimes \vert \xi_{n-1} \rangle.
\end{equation}

Using the $\mket{\Xi}$ register, you can easily show how the use of a single Hadamard gate works:
\begin{equation}
\begin{array}{c}
H\vert \Xi \rangle = H( \vert \xi_{0} \rangle \otimes \vert \xi_{1} \rangle \otimes \dots \otimes \vert \xi_{n-2} \rangle \otimes \vert \xi_{n-1} \rangle) = \\
\\
= H\vert \xi_{0} \rangle \otimes H\vert \xi_{1} \rangle \otimes \dots \otimes H\vert \xi_{n-2} \rangle \otimes H \vert \xi_{n-1} \rangle = \\
\\
= \bigotimes_{i=0}^{n-1} H \vert \xi_{i} \rangle = \frac{1}{\sqrt{2^n}} \left[ \bigotimes_{i=0}^{n-1} \big( 
\vert 0 \rangle + (-1)^{\xi_i} \vert 1 \rangle 
\big) \right]. \\
\end{array}
\end{equation}

The use of the Hadamard gateway results in the amplitude values being equal to the absolute value $\frac{1}{\sqrt{2^n}}$ for all states taken by the specified quantum register.

Should be also pay attention to the construction of controlled gates, which construction requires in addition to the tensor product the use of direct sum of matrices and also uses the projection matrix.

The one of example may be one of the possible implementations of the CNOT gate (controlled negation gate) for qubits:
\begin{equation}
U = \mket{000}\mbra{000} I + \mket{010}\mbra{010} X + \mket{111}\mbra{111}  I
\end{equation}
This gate performs a qubit negation operation, in case if the so-called control qubit takes the state one.

The second type of basic operation is the so-called measurement operation. We will present only one example of this type of operation: a von Neumann type measurement. It begins with the preparation of observables. Operator, the form of which is presented in the following way:
\begin{equation}
M = \sum_{i} \lambda_i P_i ,
\end{equation}
where $P_i$ is a projector for the operator's eigenspace $M$ with the eigenvalue $\lambda_i$.

The results of the measurement performed represent the eigenvalues $\lambda_i$. Wherein the individual results occur with following probability:
\begin{equation}
P( \lambda_i ) = \mbra{\psi} P_i \mket{\psi} .
\label{lbl:eq:prob:measure:AMCS2018:MS:MW}
\end{equation}

The obtained result $\lambda_i$ indicates that the $\mket{\psi}$ register has been transformed to:
\begin{equation}
\mket{\psi^{'}} = \frac{P_i \mket{\psi}}{\sqrt{\lambda_i}},
\end{equation}
with the probability determined by the Eq.~\ref{lbl:eq:prob:measure:AMCS2018:MS:MW}.

\section{Quantum recommendation system} \label{lbl:sec:QRS:AMCS2018:MS:MW}

This part presents the construction of a quantum recommendation system. An example of a database is indicated, from which the elements indicated by users will be recommended. The quantum register was also described in \ref{lbl:sec:DB:Structure:AMCS2018:MS:MW}. The point \ref{lbl:sec:QRS:Alg:AMCS2018:MS:MW} describes the necessary calculation steps for the implementation of the quantum recommendation algorithm.
The construction of a quantum circuit is also indicated in part \ref{lbl:ss:qcircuit:for:QRS:AMCS2018:MS:MW}. Analysis of the correctness of the proposed algorithm is given in the point \ref{lbl:sec:algebraic:correctness:AMCS2018:MS:MW}. An analytical description of the probability of success of indicating the recommended element (or many elements) has been presented.

\subsection{Database and quantum register structure} \label{lbl:sec:DB:Structure:AMCS2018:MS:MW}

\begin{figure*}
	\begin{center}
		\includegraphics[width=17cm]{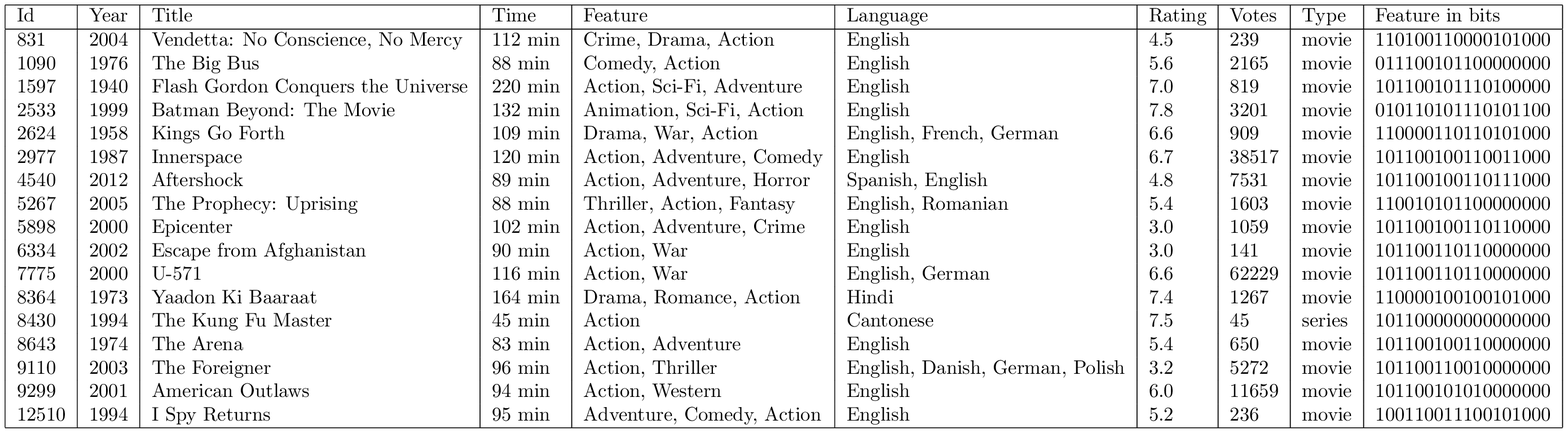}
	\end{center}
	\caption{The exemplary structure of the classic database. It should be noted that in the quantum part of the database only two columns are used: identifier "ID" and description of the feature "Feature". The indicated fields are also the main key within the classical base}
	\label{lbl:fig:fig0:classical:DB:AMCS18:MS:MW}
\end{figure*}

The proposal for building a quantum recommendation system (QRS - quantum recommendation system) presented in this article is a hybrid system. The database from which items are recommended based on the user's suggestions and needs is naturally stored in the classical system. As an example of a classical database on the basis of which a quantum recommendation system can be built, the IMDB movie database \cite{OMDB:2018}. Fig.~\ref{lbl:fig:fig0:classical:DB:AMCS18:MS:MW} shows several selected records from the database with a field describing the feature of the particular movie.

\begin{figure}
\begin{center}
\includegraphics[width=6.5cm]{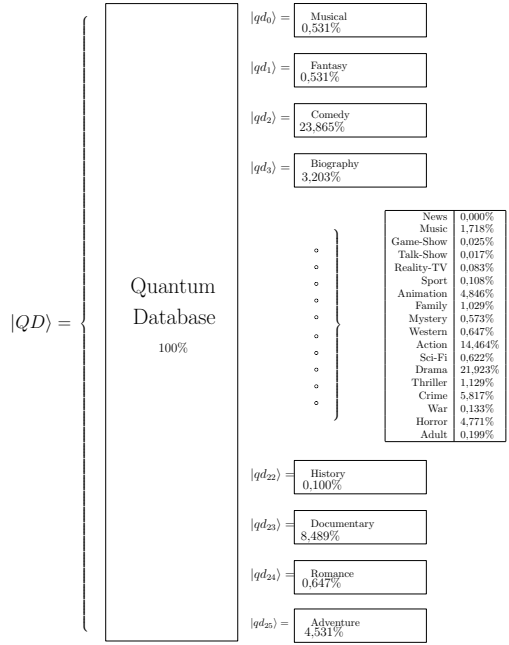}
\end{center}
\caption{Division of the main quantum registry into subregisters in the proposed recommendation system. It is assumed that individual sub-registers contain entries with the same leading feature}
\label{lbl:fig:fig1:QR:divisions:AMCS18:MS:MW}
\label{fig:qdatabase:registry1}
\end{figure}

It should be emphasized that there is no need store the entire database into the quantum system. Usually only two columns of data are relevant: identifier of recommended elements and element's feature saved as a binary word. The length of words that are used to describe the identifier and characteristics are important. It is assumed that these will be binary numbers, the identifier will be described with $q$ bits, while the feature with $l$ bits. Usually the entire of the classic database can be divided into sections referring to elements with common features. Based on this, you can create a quantum register that contains a information about the database. Representation of this register and its possible division is shown in Fig.~\ref{fig:qdatabase:registry1}. If we use the entire database in this case we have only one register, or divide the whole register into sub-registers with common features, e.g. all historical films would be collected in one register.

In view of the remarks about the form of the database and the structure of the quantum register, we distinguish four main parts:
\begin{equation}
\mket{\Psi} = \mket{\psi_q} \otimes \mket{\psi_{f_f}} \otimes \mket{\psi_{f_u}} \mket{\psi_{k}},
\end{equation}
however the order of quantum sub-register is not important, the order can be changed e.g. in the case of features: $\mket{\Psi} = \mket{\psi_q} \otimes \mket{\psi_{f_u}} \mket{\psi_{f_f}} \otimes \mket{\psi_{k}}$.

The first $q$ qubits of the registry $\mket{\Psi}$ i.e. $\mket{\psi_q}$ is used to encode the identifier of the recommended element. The next part of the $\mket{\psi_{f_f}}$ register describes the features of individual database elements. They refer naturally to the identifier of the individuals items. The $l$ qubits are used to describe the feature, because thanks to the superposition properties, we will create $2^l$ various features in the database containing $2^q$ of classical elements. The third part of $\mket{\psi_{f_u}}$ is a description of the feature desired by the user. Once again, $l$ qubits are used, although we describe only one feature desired by the user. Again by the superposition principle, applying only $l$ operations (the feature is described by a binary number with a width of $l$ bits) we connect information about user features with the rest of database. The last part denoted by $\mket{\psi_{k}}$ register represents an additional qubits. The more precise amount was specified a later at point \ref{lbl:ss:qcircuit:for:QRS:AMCS2018:MS:MW}. The point \ref{lbl:sec:algebraic:correctness:AMCS2018:MS:MW} presents an analytical analysis a probability success of the recommendation is described by $P_{max}$ quantity.

\begin{remark}{Quantity of recommendations}
In the description of the registry structure, we have provided only one feature for the user $\mket{\psi_{f_u}}$, of course, you can add additional user features, e.g. $\mket{\psi_{f_{u_1}}}, \mket{\psi_{f_{u_2}}}, \ldots, \mket{\psi_{f_{u_K}}}$, to compute a $K$ recommendation. However, in the remainder of this article, for simplicity, we limit our analysis to a single feature. $\square$
\end{remark}

\subsection{The main algorithm QRS} \label{lbl:sec:QRS:Alg:AMCS2018:MS:MW}

The scheme of proceeding in the proposed approach to the recommendation system is based on two main stages. The first one points the database elements whose features are closest to those indicated by the user. For this purpose, the quantum version of the k-nearest neighbors algorithm is used. Then, to amplify the effectiveness of the recommendation, the Grover's algorithm is used. The individual computational steps are presented as the Algorithm~\ref{lbl:alg:RCS:AMCS2018:MS:MW}.

\begin{algorithm}
	\caption{Quantum recommendation system}
	Steps of conduct in the quantum recommendation system:
	\begin{itemize}
		\item[(I)] creation of database, in this step we use $\mket{\psi_{q}}$ and $\mket{\psi_{f_f}}$ registers,
		\item[(II)] the user determines which features should have recommended elements, 
		information is represented by the register $\mket{\psi_{f_u}}$,
		\item[(III)] for a given feature, the appropriate sub-register (or a whole quantum register) is selected representing the relevant part of the database, data are encoded in the state of register $\mket{\psi_{f_f}}$,
		\item[(IV)] a recommendation process is performed using a quantum algorithm of k-nearest neighbours, two register are used $\mket{\psi_{f_u}}$ and $\mket{\psi_{f_f}}$, we also use auxiliary qubit from $\mket{\psi_{k}}$ called $\mket{c_0}$, 
		\item[(Va)] if the state of qubit $\mket{c_0}$ after measurement operation is $\mket{0}$, then the obtained probability distribution of the recommended elements from step (IV) can be amplified by Grover's algorithm to improve the probabilistic properties of the best recommended elements, only state of register $\mket{\psi_{f_u}}$ is amplifying. We can jump to step (VI),
		\item[(Vb)] If the state of qubit $\mket{c_0}$ after measurement operation is $\mket{1}$, then we uncompute q-kNN part and the step (IV) must be repeated,
		\item[(VI)]performing the second measurement on the quantum register representing the database, finally the recommended element will be obtained (i.e. the element that will have the highest probability of measurement).
	\end{itemize}
	\label{lbl:alg:RCS:AMCS2018:MS:MW}
\end{algorithm}

\begin{figure}
	\begin{center}
		\includegraphics[width=5.00cm]{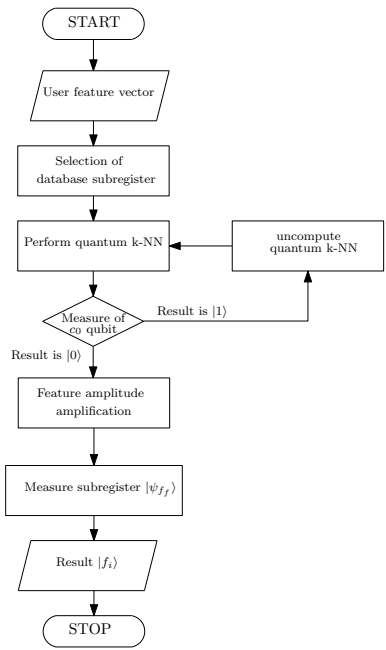}
	\end{center}	
	\caption{The control flow of proposed the quantum recomendation system}
	\label{fig:control:flow:AMCS18:MS:MW}
\end{figure}

The algorithm~\ref{lbl:alg:RCS:AMCS2018:MS:MW} can also be represented as a control flow in the proposed approach to the recommendation system. The corresponding diagram is shown in Fig.~\ref{fig:control:flow:AMCS18:MS:MW}. Because the algorithm is based on measurement operations, an additional qubit called $c_0$ has been introduced in the quantum register. Performing the measurement operation on the qubit $c_0$, allows you to determine whether you have successfully converted the registry to the correct probability distribution. The measurement result $c_0$ assuming the state $\mket{0}$ means that the quantum register has been transformed into a proper state, and it is probability distribution for the next measurement will indicate a recommended element.

\subsection{Quantum circuit for QRS and computational complexity} \label{lbl:ss:qcircuit:for:QRS:AMCS2018:MS:MW}

\begin{figure*}
\begin{center}
	\includegraphics[width=17cm]{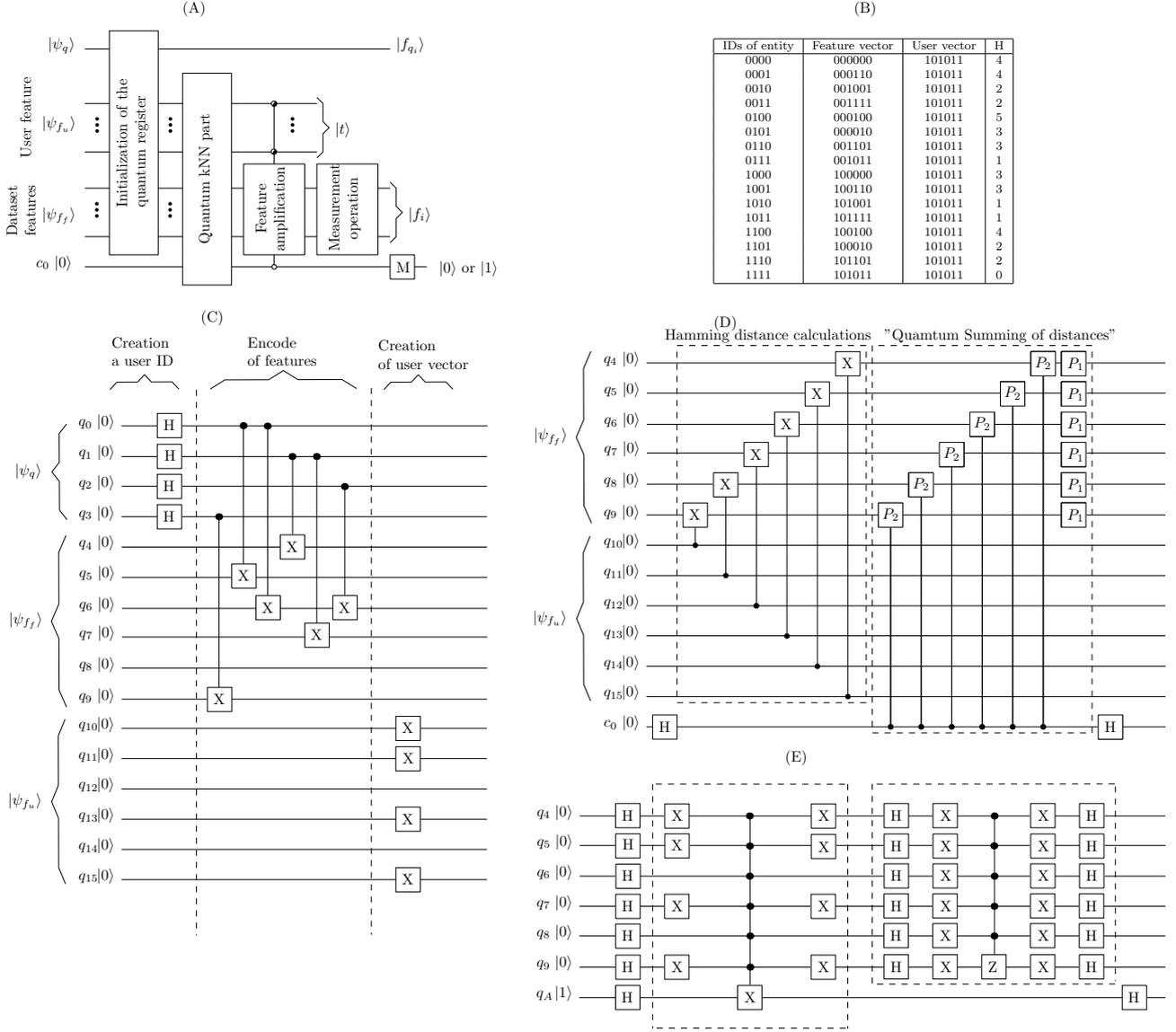}
\end{center}	
	\caption{The diagram of the quantum circuit for discussed recommendation system. The diagram (A) presents three main elements of the recommendation process: preparation of the database, implementation of a quantum algorithm and amplify the recommended element. The table (B) presents a system of recommended items that were used to construct exemplary circuits (the column H shows the Hamming distance between element and example feature $101011$ defined by the user). The quantum circuit (C) represents the initialization part of the database. Diagram (D) shows an example of the implementation of the main part of quantum k-NN algorithm: Hamming distance calculation and so-called quantum summing of distance. The last diagram (E) is a part of amplified amplitudes for the user's feature, i.e. $101011$}
	\label{fig:QPD:general:circuit}
\end{figure*}

Fig.~\ref{fig:QPD:general:circuit} part (A) shows the general structure of the quantum circuit implementing the quantum recommendation system. Three exemplary circuits implementing the most important stages of the recommendation system were also given there. The examples are based on a 16-element database shown in table (B) on Fig.~\ref{fig:QPD:general:circuit}.

Diagram (C) in the Fig.~\ref{fig:QPD:general:circuit} shows the database initialization process, which consists of three stages. The first stage creates database identifiers, for this purpose it applies exactly $l$ Hadamard gates. The next stage is the encoding of features, unfortunately it requires a larger amount of operation which depends on the number of elements in the database. In general, after creating identifiers in the $\mket{\psi_q}$ register, the permutation of elements must be performed in the register $\mket{\psi_{f_f}}$. Permutation will generally requires no more than $N(N-1)/2$ single qubit and controlled gates \cite{Li2013}, with $N=2^q$. Unfortunately, the process of creating a database will not be significantly better than the creation of a classic database, although the structure of feature codes may be helpful in the process of creating a database. It can also be seen in the presented example, where six operators of controlled negation are enough (the given relation defines the upper limit, therefore for 16 elements, the maximum number of gates is 120) to generate 16 database entries. The third stage, due to superposition principle will require only $l$ negation gates to build the user's feature. Computational complexity in the $O$ notation of this stage can be written as:
\begin{equation}
O_1(l,N)=2l + N(N-1)/2 .
\end{equation}

However, as shown in the correctness analysis \ref{lbl:sec:algebraic:correctness:AMCS2018:MS:MW}, you do not need to repeat the process of building the entire database. Circuit (D) presents the process of calculating the Hamming distance and the so-called sum of distances to make a classification. The recommendation process is completed after the measurement has been made auxiliary qubits $c_0$, if the result measurement is $\mket{0}$, further details are presented in sec.~\ref{lbl:sec:algebraic:correctness:AMCS2018:MS:MW}. The number of operations to be performed when calculating the Hamming distance (D) is linear and depends on the width of the feature, i.e. the number of classic bits $l$ describing the feature:
\begin{equation}
O_2(l) = 3l + 2 .
\end{equation}

\begin{remark}{Complexity of q-kNN part}
Particular attention should be paid to the fact that the computational complexity of this step depends only on the width of the pattern. Unlike to the classic methods, where the number of rows should be taken into account for the process to analysis of computational complexity.
$\square$
\end{remark}

The construction of the amplifying amplitude circuit for one or several amplitudes with the Grover algorithm depends on the form of the feature described by the user. Plot (E) in Fig.~\ref{fig:QPD:general:circuit} shows the amplification of one amplitude for exemplary database shown in table (B) in Fig.~\ref{fig:QPD:general:circuit}. The number and arrangement of the negation gates in the oracle section is an exact mapping of the feature specified by the user. Definition of oracle can be simplified by using only negation on the bits that are encoded, which encode the main feature, e.g. a historical movie, can be encoded with the oldest or youngest bit. However, it can be assumed that the feature requires the use of the largest number of gates, their number can be defined as:

\begin{equation}
O_3(l) =  7l + 2c + 3.
\end{equation}
where $l$ is the width of the feature, and $c$ means the number of gates after decomposing the negation gate in stage of the oracle of the Grover algorithm and in the controlled gate $Z$ in the part performing the rotation around the average. Decomposition of controlled gates can be done according to the works \cite{Barenco1995} and \cite{Shende:2009}, however, and because we have $l$ wide feature then finally we can obtain polynomial complexity (in addition we still operate on $2^q$-th classical data).

All computational complexity is naturally dominated by the database creation process, but the calculation process of the recommendation depends only on the width of the feature, and not on the amount of data in the database.

\subsection{Correctness of quantum recommendation system} \label{lbl:sec:algebraic:correctness:AMCS2018:MS:MW}

Algebraic correctness  of quantum recommendation system will be started by defining the initial state of a quantum register during the first stage, i.e. preparation of the database (see ). For easy further analysis, qubits describing the record identifier will be omitted:
\begin{equation}
\vert \psi_0 \rangle = \vert 0 \rangle^{\otimes 2l+1}.
\end{equation}
The first $l$ qubits represents the database file, the second $l$ qubits, the user feature vector. Initializing the sub-register with the feature database results in obtaining the following status:
\begin{equation}
\mket{\Psi_1} = \frac{1}{\sqrt{L}} \sum_{p=1}^{L}\vert r_{1}^{p}, \dots, r_{l}^{p} \rangle,
\label{lbl:eq:initial:db:state}
\end{equation}
where $L=2^l$, and $r_p^{k}$ is a bit description of the features a selected row in the database. It is obvious that, for transparency the part $\mket{\psi_q}$ was omitted from Fig.\ref{fig:qdatabase:registry1}.

After entering the form of the user's feature vector, the state of the quantum register is given the form:
\begin{equation}
\mket{\Psi_2} = \frac{1}{\sqrt{L}} \left( \sum_{p=1}^{L}\vert r_{1}^{p}, \dots, r_{l}^{p} \rangle \right) \otimes \vert t_{1}, \dots, t_{l} \rangle \otimes \frac{1}{\sqrt{2}}(\vert 0 \rangle + \vert 1 \rangle) .
\end{equation}
From that moment, we can calculate the distance between Hamming between the user's characteristics and the features stored in the database table. Responsible for distance calculation in quantum circuit (D) on Fig.~\ref{fig:QPD:general:circuit} is first set of controlled gates NOT which is described by unitary operations $U$:
\begin{equation}
U=e^{-\mathbf{i}\frac{\pi}{2l}\hat{H}}, \;\;\; \hat{H}= \left[
\begin{array}{cc}
1 & 0\\
0 & 0\\
\end{array}
\right]^{\otimes l} \\ \otimes I_{{l} \times {l}} \otimes \left[
\begin{array}{cc}
1 & 0\\
0 & -1\\
\end{array}
\right] .
\label{lbl:eq:U:operation}
\end{equation}

The operation $U$ can be decomposed on the set of single qubit gates $P_1$:
\begin{equation}
P_1 = \left[
\begin{array}{cc}
e^{-\mathbf{i}\frac{\pi}{2l}} & 0\\
0 & 1\\
\end{array}
\right],
\end{equation}
and controlled gate:
\begin{equation}
P_2 =  \mket{0}\mbra{0} + \mket{1}\mbra{1} \otimes {P_1}^{-2}.
\end{equation}
Additional discussion about decomposition of $U$ can be found at work \cite{Trugenberger:2002}.

The use of the $ U $ operation described by equation Eq.(\ref{lbl:eq:U:operation}) causes that we get the following state:
\begin{equation}
\vert \psi_4 \rangle = \frac{1}{\sqrt{L}} \sum_{p=1}^{L} \left( e^{\mathbf{i}\frac{\pi}{2l}d(t,r^p)} \vert d_{1}^{p}, \dots, d_{l}^{p} \rangle \right. \\
\left. \otimes \vert t_{1}, \dots, t_{l} \rangle \otimes \vert 0 \rangle + e^{-\mathbf{i}\frac{\pi}{2l}d(t,r^p)} \vert d_{1}^{p}, \dots, d_{l}1^{p} \rangle \right. \\
\left. \otimes \vert t_{1}, \dots, t_{l} \rangle \otimes \vert 1 \rangle \right) .
\end{equation}
The perform of Hadamard operation on the additional qubit $c_0$ allows to obtain the final state for the stage related to the quantum k-NN algorithm:
\begin{multline}
\mket{ \psi_5 } = \frac{1}{\sqrt{L}} \sum_{p=1}^{L} \left( \cos \left(\frac{\pi}{2l}d(t,r^p) \right) \vert d_{1}^{p}, \dots, d_{l}^{p} \rangle \right. \\
\left. \otimes  \vert t_{1}, \dots, t_{l} \rangle \otimes \vert 0 \rangle + \sin \left( \frac{\pi}{2l}d(t,r^p) \right)  \vert d_{1}^{p}, \dots, d_{l}^{p} \rangle \right. \\
\left. \otimes \vert t_{1}, \dots, t_{l} \rangle \otimes \vert 1 \rangle \right).
\label{lbl:eq:final:state}
\end{multline}	
The probability of measuring zero on the state $c_0$, i.e. success, that we will go to the desired probability distribution with the correct indication of the recommended element:
\begin{equation}
P(c_0) = \frac{1}{L} \sum_{p} \cos^2 \left(\frac{\pi}{2l}d(t,r^p) \right).
\label{lbl:eq:pc0:prob}
\end{equation}

If you have measured the state $\mket{1}$ on qubits $c_0$, then you should restore the state $\mket{\psi_4}$. You can do this by performing an inverse operation to $U$ given by Eq.~\ref{lbl:eq:U:operation} i.e. $U^{\dagger}$. This operation will require the same amount of work as the $U$ operation and once again use the negated control operators, which are self-adjoint, which will restore the state of the registry to the state before Hamming distance calculation and summing. In this way, we avoid the costly process of rebuilding the database.

If, by simplifying, by $P$ we denote the probability of receiving a particular recommendation, then obtaining a probability distribution with accuracy of $\varepsilon$ requires in generally $O(P \cdot (1 - P) \cdot  \frac{1}{\varepsilon^2} )$ repetitions of the implementation a quantum algorithm of the k-nearest neighbors. It should be emphasized, however, that using classical multiprocessor solutions, you can use many quantum machines to solve the same task, for example to obtain a linear time calculation of the probability distribution for the exponential amount of data $L$, described by Eq.~(\ref{lbl:eq:initial:db:state}).

\begin{remark}{Improving the distribution of the probability of success}
The distribution of the probability obtained is not satisfactory if the user's designated feature determines the choice of one compatible item or a few from a very close neighborhood. Therefore, using the Grover algorithm significantly improves the final probability distribution of the received recommendation \cite{Brassard:1997}. $\square$
\end{remark}
Therefore, assuming that zero was obtained by measuring the state of $c_0$, we get the state of:
\begin{equation}
\mket{\psi_6} = \frac{1}{\sqrt{L}} \sum_{p=1}^{L} m_p  \mket{\psi_{rcmd}} ,
\end{equation}	
wherein $\mket{\psi_{rcmd}} = \vert d_{1}^{p}, \dots, d_{l}^{p} \rangle \otimes  \vert t_{1}, \dots, t_{l} \rangle \otimes \mket{0}$, whereas $m_p$ is represented by the probability amplitudes obtained after the measurement. In the existing registry, you can emphasize $g$ amplitudes for recommended elements with the highest compatibility value a feature of the $m^r_p$ and also $L-g$ with lower compatibility $m^{nr}_p$:
\begin{equation}
\mket{\psi_7} = \frac{1}{\sqrt{L}} \left( \sum_{p=1}^{g} m^r_p \mket{\psi_{rcmd}} + \sum_{p=g+1}^{L} m^{nr}_p \mket{\psi_{rcmd}} \right) .
\end{equation}
Average for amplitudes and variance for probability amplitudes for state $\mket{\psi_7}$, takes the following form based on the article \cite{Biham:1999}:
\begin{equation}
\begin{array}{c}
\overline{m^r} (t) = \frac{1}{g} \sum_{p=1}^{g} m^r_p ( t ), \\ \\
\overline{m^{nr}} (t) = \frac{1}{g} \sum_{p=g+1}^{L} m^{nr}_p ( t ) , \\ \\
\sigma^2_{nr}(t) = \frac{1}{L-g}  \sum_{p=g+1}^{L} | m^{nr}_p ( t ) - \overline{m^{nr}} (t) |^2,
\end{array}
\end{equation}
where $t$ is the iteration number commonly referred to as the execution time of the Grover algorithm, for $t=0$, of course, the initial values of the distribution of probability amplitudes are known \cite{Biham:1999}. The highest probability of measuring the existing recommended elements from the database is described as follows:
\begin{multline}
P_{max} = 1 - (L-g) \sigma^2_{nr} - \frac{1}{2} \left( (L-g) {|\overline{m^{nr}}(0)|}^2 \right.  \\ 
\left. + g  {|\overline{m^r} (0)|}^2 \right) + \left( \frac{1}{2} | (L-g) {\overline{m^{nr}}(0)}^2  + g  {\overline{m^r} (0)}^2 | \right),
\label{lbl:eq:P:max}
\end{multline}
with $O(\sqrt{\frac{L}{g\textbf{}}})$ iteration.

\section{Numerical experiments} \label{lbl:sec:numerical:experiment:AMCS2018:MS:MW}

\begin{figure*}
	\begin{center}
		\begin{tabular}{ccc}
			(A) & (B) & (C) \\
			\includegraphics[width=5.25cm]{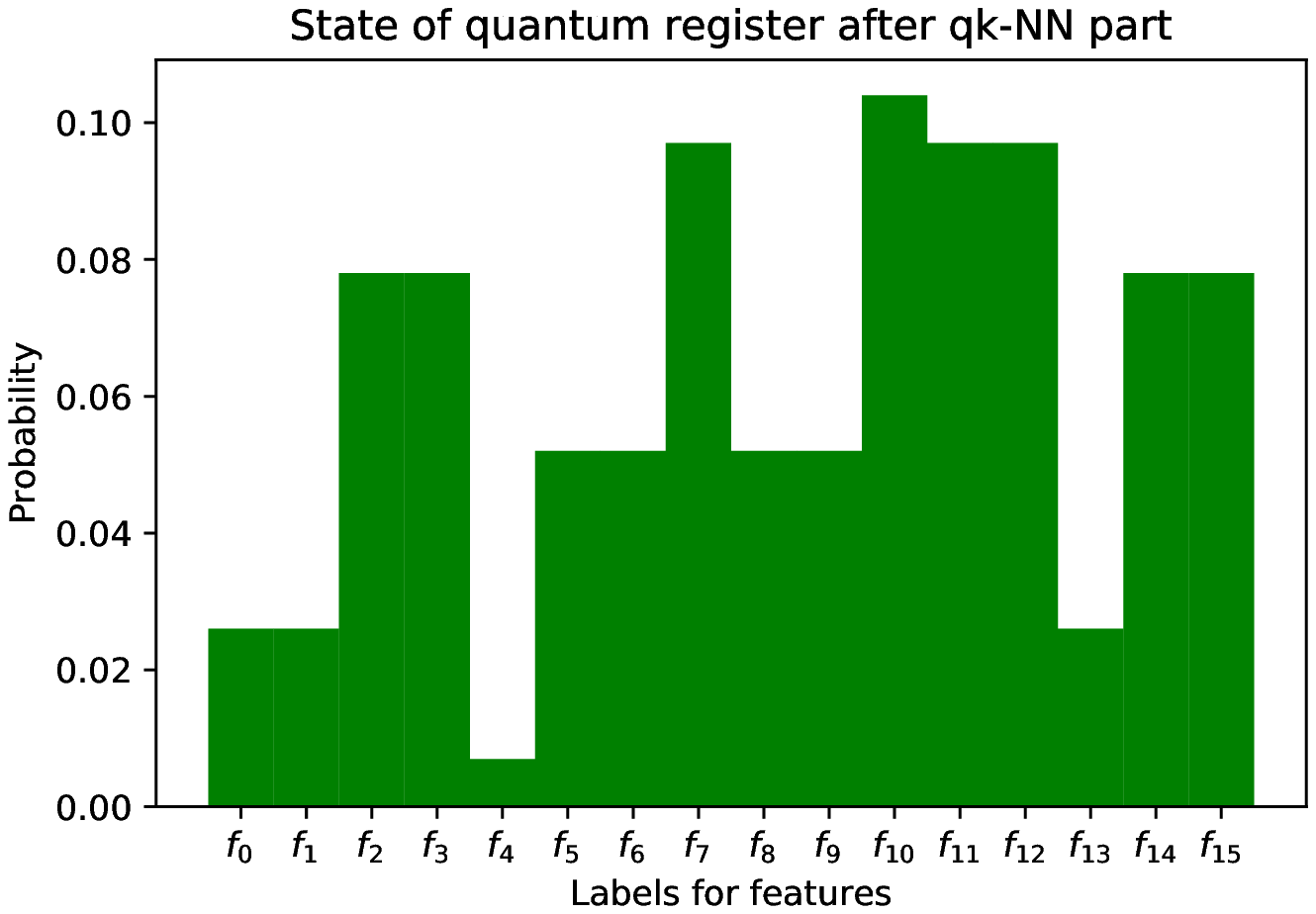} & \includegraphics[width=5.25cm]{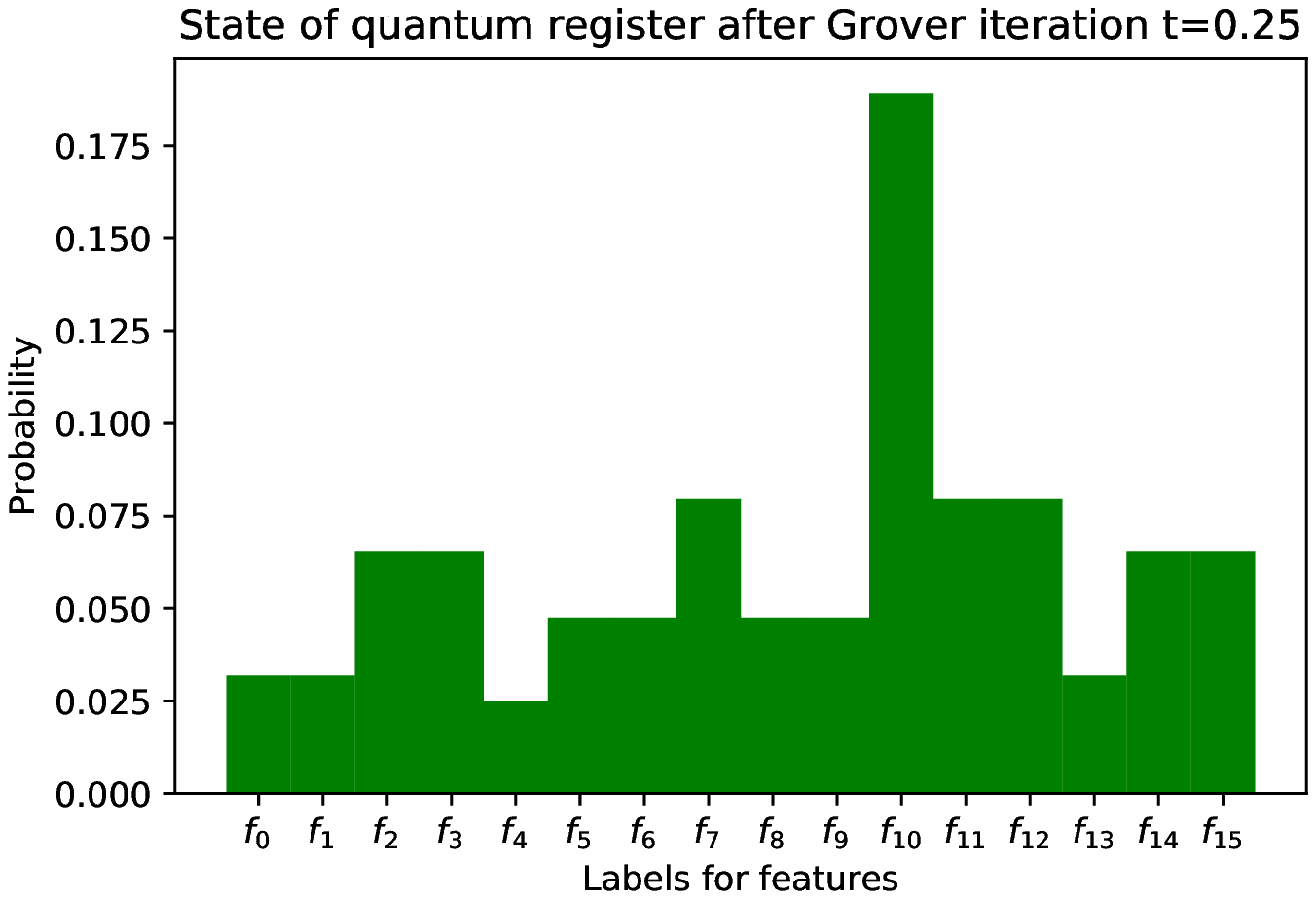} & \includegraphics[width=5.25cm]{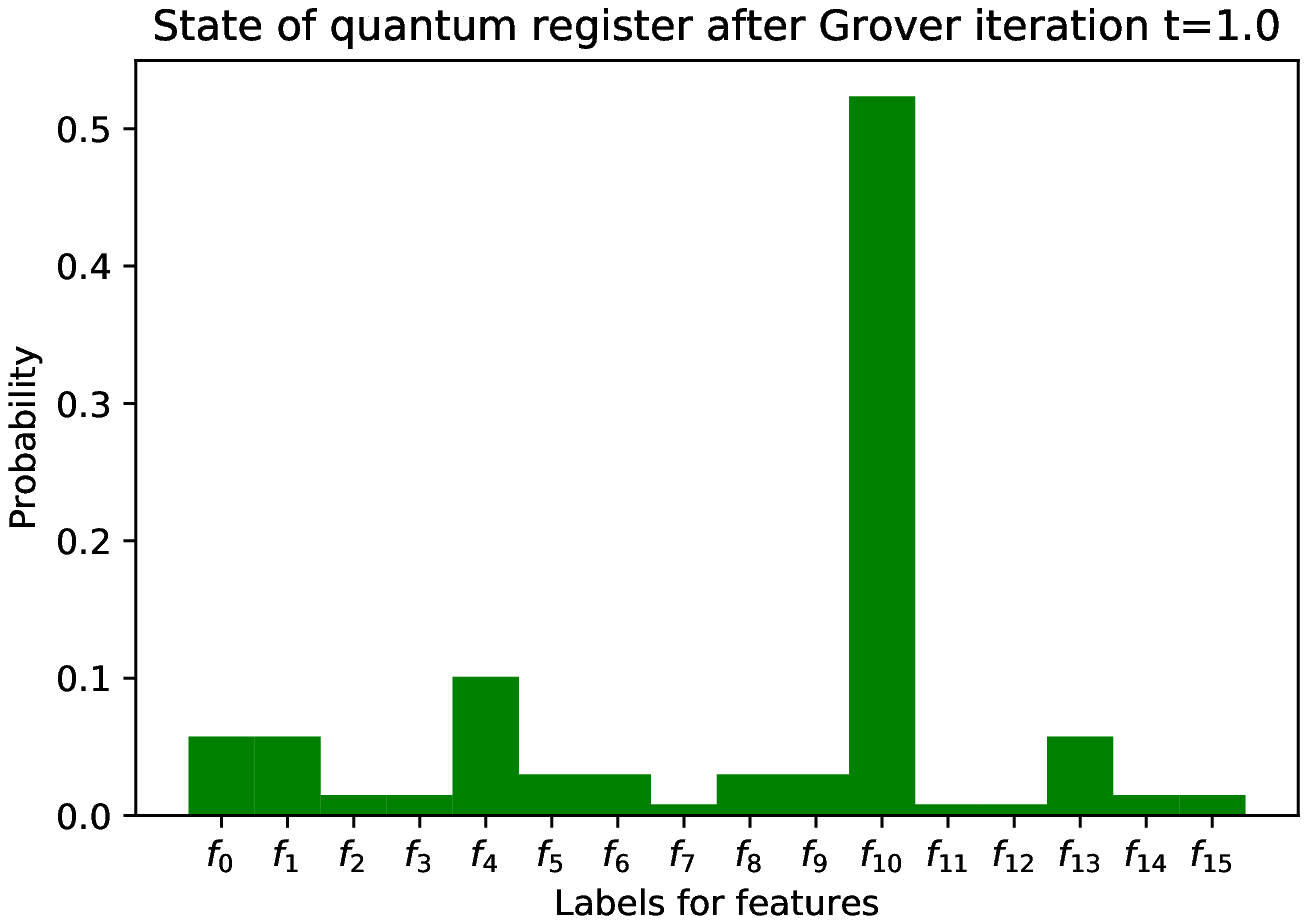} \\
			(D) & (E) & (F) \\
			\includegraphics[width=5.25cm]{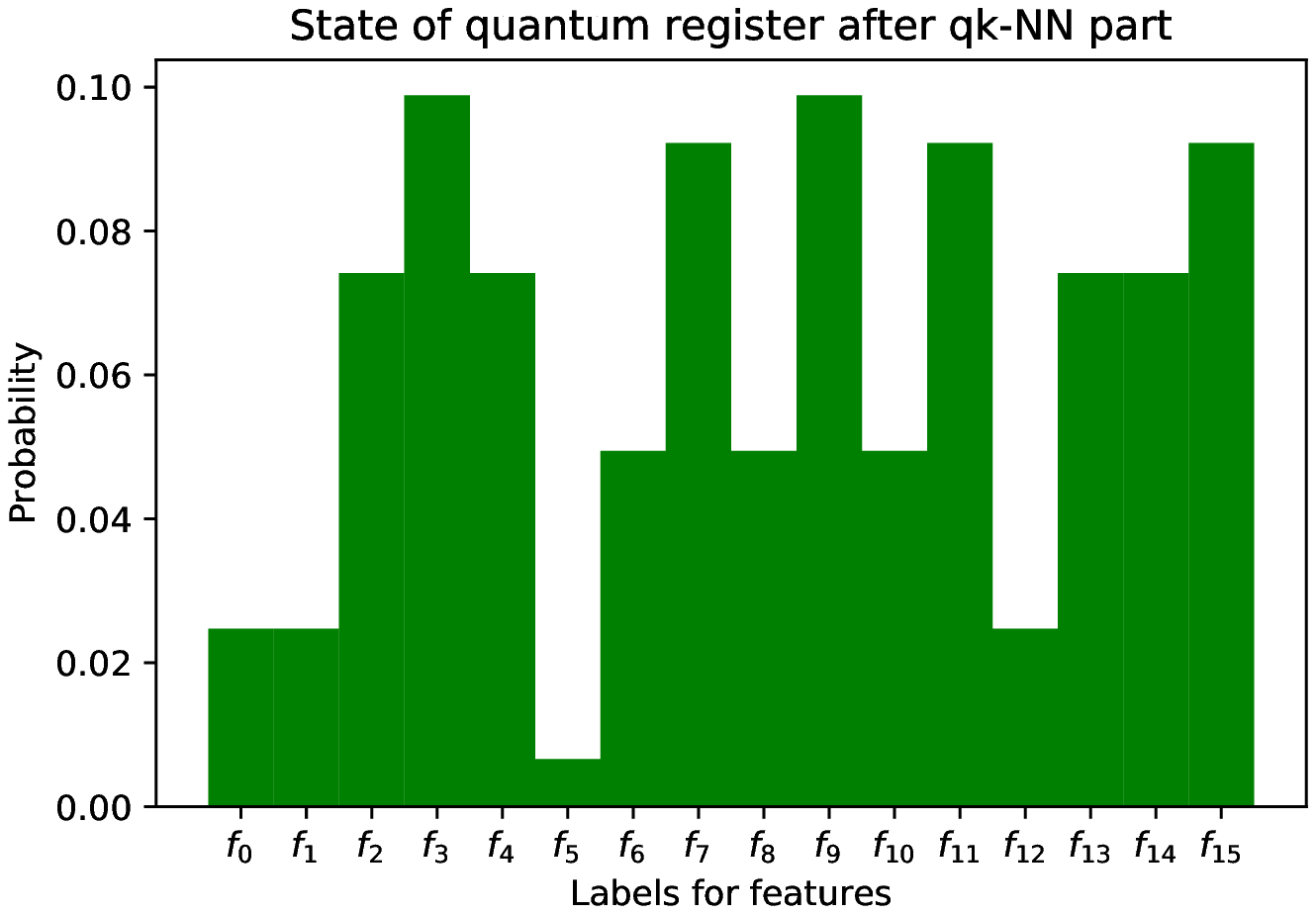} & \includegraphics[width=5.25cm]{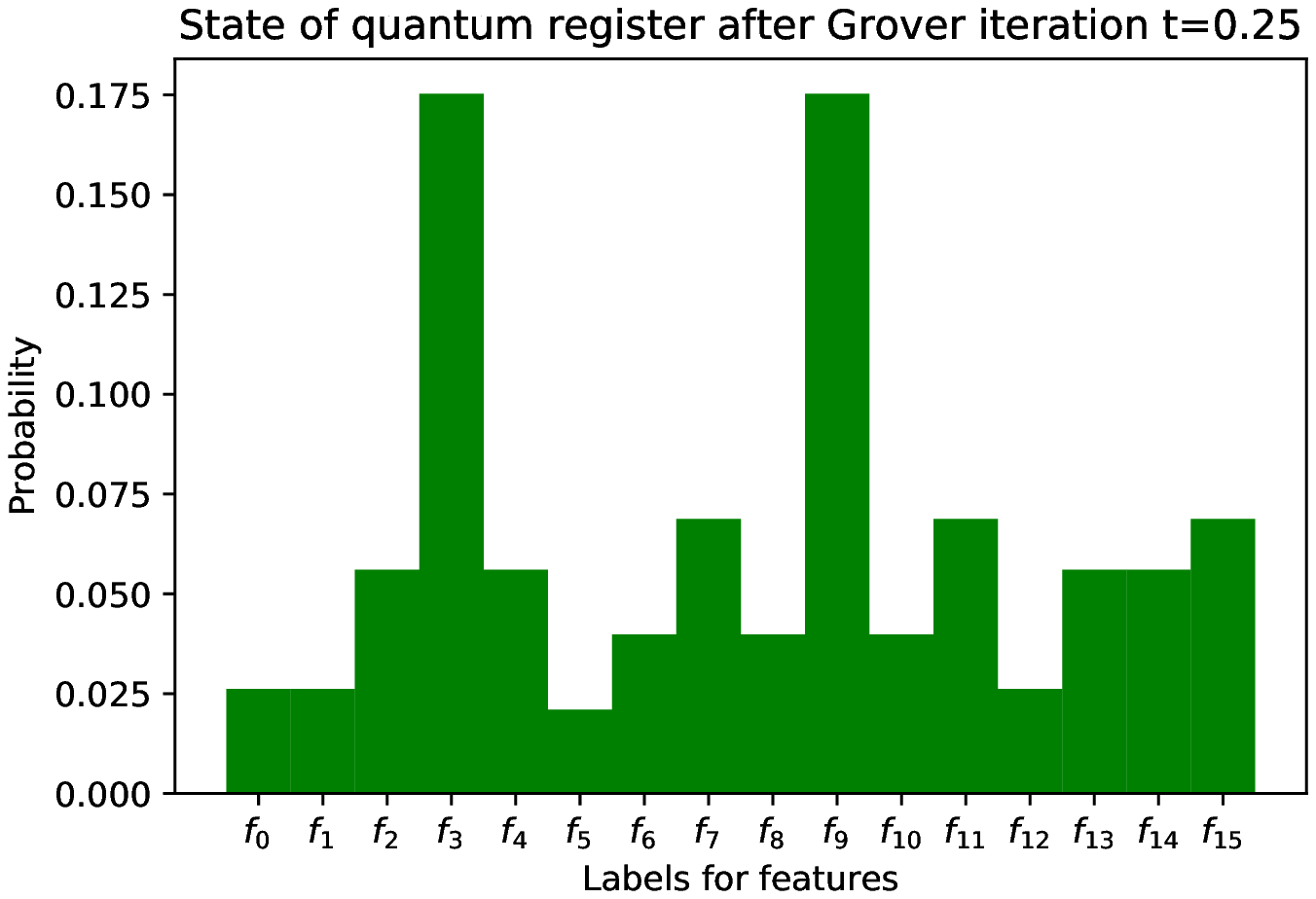} & \includegraphics[width=5.25cm]{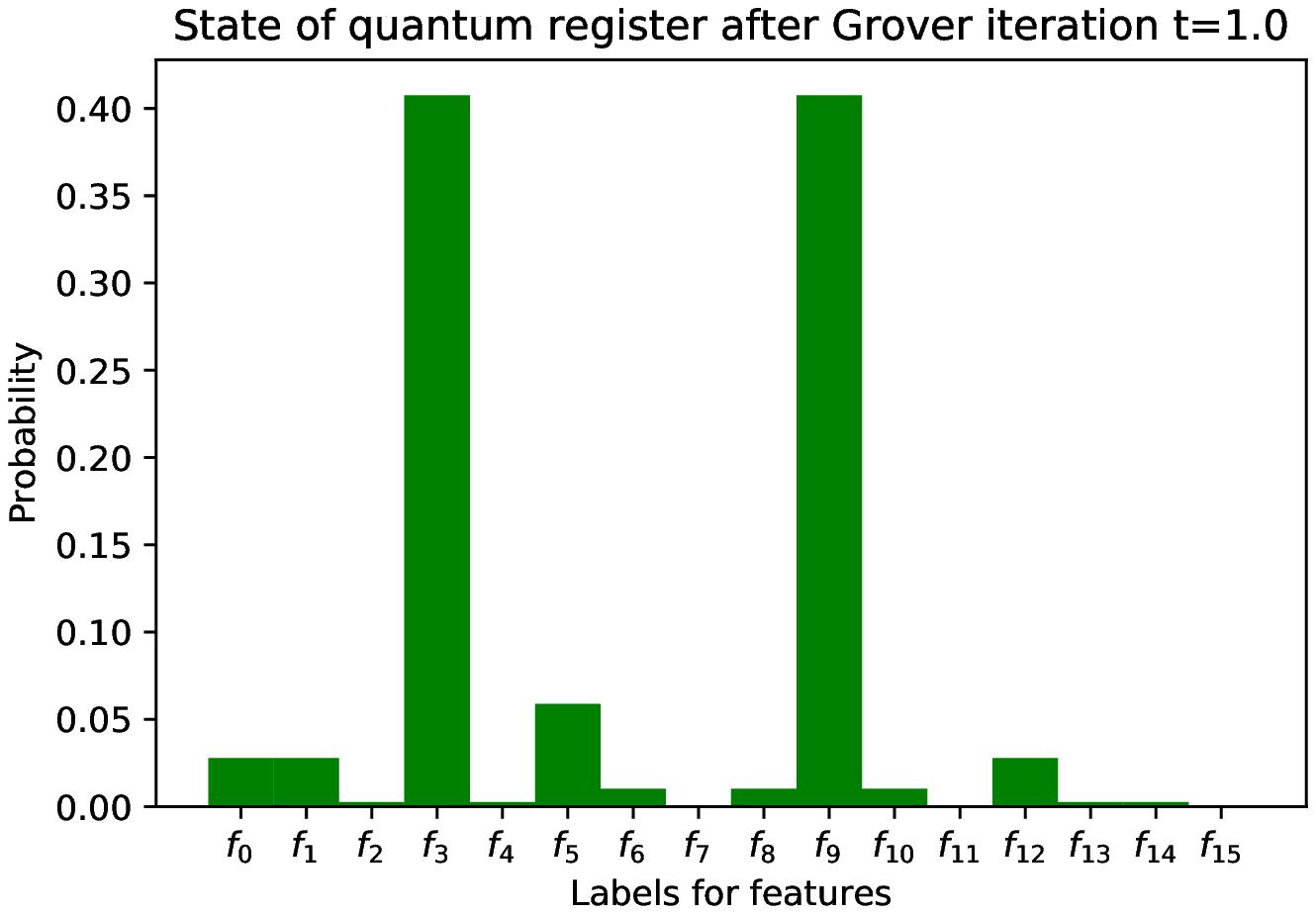} \\
		\end{tabular}
	\end{center}
    \caption{The state of the quantum register (for a database with 16 elements) in three main stages in the proposed recommendation system. The probabilty after initial datasbase is not shown. In such case every feature $f_i$ has amplitude probability equal to 1/16. Cases (A) and (D) represents the quantum state after the execution of the part realisations the k-NN algorithm. Cases (B), (C) and (E), (F) show the process of amplifying the amplitude of one (feature $f_{10}$) and two recommended elements ($f_3$, $f_9$) in the database using the Grover algorithm.}
\label{lbl:fig:database:numerical:recomd:example}
\end{figure*}

The given algebraic relations describing the states of the quantum register after particular stages of the implementation of Algorithm~\ref{lbl:alg:RCS:AMCS2018:MS:MW}  allow us to give numerical examples presenting the behaviour of the recommending system. Fig.~\ref{lbl:fig:database:numerical:recomd:example} presents probability distributions showing the state of the registry for a database of sixteen elements. There have been cases where one $f_ {15}$ amplified in the database or two $f_ {6}$ and $f_ {13}$ elements that match the user's expectations.

Regardless of the case, the plots (A) and (D) on Fig.~\ref{lbl:fig:database:numerical:recomd:example}, shows the probability distribution for the database after the execution of the k-NN part. In both cases (Fig.~\ref{lbl:fig:database:numerical:recomd:example} graphs (A) and (S)) we have no explicit indication of the element nearest to the user feature. Although, for the eligible feature, for example $f_{10}$, we have the highest probability of measuring this state. However, due to the fact that the difference in Hamming distance for the other features was not significantly large, the other elements also have a high probability of measurement. If we stop the process of recommendation at this stage, we should repeat the implementation of steps (I), (II), (III), (IV) from Alg.~\ref{lbl:alg:RCS:AMCS2018:MS:MW} and the measurement operation to collect the results which the probability described by Eq.~\ref{lbl:eq:pc0:prob}.

It should also be emphasized that in charts (B), (C) we describe the case of a recommendation of only one element. However, the quality of the amplification will be good for two or more elements that will be indicated by a user-specified characteristic. What is presented in graphs (E), (F) of Fig.\ref{lbl:fig:database:numerical:recomd:example}. It should be emphasized that the sum of probabilities of recommended elements is determined by Eq.~\ref{lbl:eq:P:max}.

\begin{remark}{The use of Grover's algorithm}
At this moment, you can ask whether instead of using k-NN you can use only the Grover algorithm itself to amplified the recommended element or more elements. The answer is affirmative only if there is an element or elements with this same vector of features as the user's requirements. If, however, there is no such element, then the Grover algorithm will not be able to amplify elements similar to those defined by the user. It should be added that you do not have to specify that we are interested in amplify only one element that fully complies with the user-defined characteristic. You can specify that we amplify the k-NN result for a leading user-specified characteristic (e.g. historical films, and the idea of such a gain is shown as cases (F), (G), (H) from Fig.~\ref{lbl:fig:database:numerical:recomd:example}). Therefore, the combination of two quantum k-NN algorithms and amplitude amplification using the Grover algorithm allows to building a recommendation system. $\square$
\label{lbl:rmk:App:Grover:AMCS2018:MS:MW}
\end{remark}

According to the note~\ref{lbl:rmk:App:Grover:AMCS2018:MS:MW} it is reasonable to use the Grover algorithm to amplify the probability for the element (elements) indicated by k-NN. For values for $t>1$ we already have a very high probability (close to the maximum theoretical value) to measure the recommended element. However, even for smaller values of $t$ it amplify the desired effect, while maintaining the probability distribution obtained after the quantum part k-NN. It also means shortening the circuit's operating time, because you do not have to do an additional iterations during the amplify with the help of the Grover algorithm.

\section{Conclusions} \label{lbl:sec:conclusions:AMCS2018:MS:MW}

The article presents the structure of the recommendation system based on the quantum k-NN algorithm and the Grover algorithm. The discussed approach is characterized by better computational complexity in the recommendation process. However, the construction of the database can be done only once, at the beginning of the mentioned stage. The process of its construction depends only on the amount of data. The probability value of the recommendation's success was also indicated. Showing that it depends directly on Hamming distance and amplitude amplification using the Grover algorithm.

One of the next tasks may be a different approach to verifying the correctness operation of the system using the quantum predicate system \cite{dhondt_panangaden_2006}, \cite{GielerakSawerwain2010}. An attempt to implement the presented system is also planned to check the quality of recommendations on existing systems of experimental installations of quantum computing systems \cite{IBM:2018}. They offer access to a quantum register of 20 qubits. This is not a sufficient number to build a larger system. However, technological advances seem to soon allow for the construction of a system containing over 50 qubits, and this size is already available in the system \cite{IBM:2018}.

The exemplary full database includes over 12,000 records. However, in the case of quantum calculations, the quantum register stores an exponential amount of data. Therefore, in the case of the OMDB movie database, indexes of individual films, we only need 15 qubits: $2^{15}>12,000$. Therefore, if nearby quantum processing systems offer access to 50 qubits, this number will allow full implementation of the proposed solution. $\square$ 

\section*{Acknowledgment}
We would like to thank for useful discussions with the~\textit{Q-INFO} group at the Institute of Control and Computation Engineering (ISSI) of the University of Zielona G\'ora, Poland. We would like also to thank to anonymous referees for useful comments on the preliminary version of this chapter. The numerical results were done using the hardware and software available at the ''GPU $\mu$-Lab'' located at the Institute of Control and Computation Engineering of the University of Zielona G\'ora, Poland. 

\bibliographystyle{plain}

\begin{thebibliography}{10}
	
	\bibitem{IBM:2018}
	{IBM} {Q} {Homepage}, 2018.
	\newblock https://quantumexperience.ng.bluemix.net/, last accessed 2018/04/28.
	
	\bibitem{OMDB:2018}
	{OMDb} {Homepag}e, 2018.
	\newblock http://www.omdbapi.com/, last accessed 2018/04/21.
	
	\bibitem{Aaronson2004}
	Scott Aaronson and Daniel Gottesman.
	\newblock Improved simulation of stabilizer circuits.
	\newblock {\em Phys. Rev. A}, 70:052328, Nov 2004.
	
	\bibitem{Alpaydin:2004}
	E.~Alpaydin.
	\newblock {\em Introduction to Machine Learning (Adaptive Computation and
		Machine Learning)}.
	\newblock The Massachusetts Institute of Technology Press, Massachusetts, 2004.
	
	\bibitem{Armbrust:2010}
	M.~Armbrust, A.~Fox, R.~Griffith, A.~D. Joseph, R.~Katz, A.~Konwinski, G.~Lee,
	D.~Patterson, A.~Rabkin, I.~Stoica, and M.~Zaharia.
	\newblock A view of cloud computing.
	\newblock {\em Communications of the Association for Computing Machinery},
	53(4):50--58, 2010.
	\newblock DOI: 10.1145/1721654.1721672.
	
	\bibitem{Barenco1995}
	Adriano Barenco, Charles~H. Bennett, Richard Cleve, David~P. DiVincenzo, Norman
	Margolus, Peter Shor, Tycho Sleator, John~A. Smolin, and Harald Weinfurter.
	\newblock Elementary gates for quantum computation.
	\newblock {\em Phys. Rev. A}, 52:3457--3467, Nov 1995.
	\newblock DOI: 10.1103/PhysRevA.52.3457.
	
	\bibitem{Biham:1999}
	E.~Biham, O.~Biham, D.~Biron, M.~Grassl, and D.~Lidar.
	\newblock Grover's quantum search algorithm for an arbitrary initial amplitude
	distribution.
	\newblock {\em Physical Review}, 60(4):2742--2745, 1999.
	\newblock DOI: 10.1103/PhysRevA.60.2742.
	
	\bibitem{Brassard:1997}
	G.~Brassard and P.~Hoyer.
	\newblock An exact quantum polynomial-time algorithm for simon's problem.
	\newblock In {\em Proceedings of the Fifth Israeli Symposium on Theory of
		Computing and Systems}, 1997.
	\newblock DOI: 10.1109/ISTCS.1997.595153.
	
	\bibitem{Busemeyer:2012}
	J.~Busemeyer and P.~Bruza.
	\newblock {\em Quantum Models of Cognition and Decision}.
	\newblock Cambridge University Press, Cambridge, 2012.
	
	\bibitem{dhondt_panangaden_2006}
	Ellie D'Hondt and Prakash Panangaden.
	\newblock Quantum weakest preconditions.
	\newblock {\em Mathematical Structures in Computer Science}, 16(3):429–451,
	2006.
	
	\bibitem{Erdal:2005}
	A.~Erdal.
	\newblock An information-theoretic analysis of grover's algorithm.
	\newblock In {\em Quantum Communication and Information Technologies}, 2003.
	\newblock ISBN: 978-94-010-0171-7.
	
	\bibitem{GielerakSawerwain2010}
	R.~Gielerak and M.~Sawerwain.
	\newblock Generalised quantum weakest preconditions.
	\newblock {\em Quantum Information Processing}, 9(4):441--449, Aug 2010.
	
	\bibitem{Grover:1996}
	L.~K. Grover.
	\newblock A fast quantum mechanical algorithm for database search.
	\newblock In {\em ANNUAL ACM SYMPOSIUM ON THEORY OF COMPUTING}, 1996.
	
	\bibitem{Schliep:2004}
	K.~Hechenbichler and K.~Schliep.
	\newblock Weighted k-nearest-neighbor techniques and ordinal classification,
	2004.
	
	\bibitem{Li2013}
	Chi-Kwong Li, Rebecca Roberts, and Xiaoyan Yin.
	\newblock Decomposition of unitary matrices and quantum gates.
	\newblock {\em International Journal of Quantum Information}, 11(01):1350015,
	2013.
	
	\bibitem{Nielsen:2010}
	M.~Nielsen and I.~Chuang.
	\newblock {\em Quantum Computation and Quantum Information: 10th Anniversary
		Edition}.
	\newblock Cambridge University Press, Cambridge, 2010.
	
	\bibitem{Nielsen:2016}
	P.~Nielsen.
	\newblock Big data analytics – a brief research synthesis.
	\newblock In {\em Information Systems Architecture and Technology}, 2016.
	\newblock ISBN: 978-3-319-28553-5.
	
	\bibitem{Pinkse:2013}
	P.W.H. Pinkse, S.A. Goorden, M.~Horstmann, B.~Skoric, and A.P. Mosk.
	\newblock Quantum pattern recognition.
	\newblock In {\em Lasers and Electro-Optics Europe (CLEO EUROPE/IQEC),
		Conference on and International Quantum Electronics Conference, Munich},
	2013.
	
	\bibitem{Sawerwain:2019}
	M.~Sawerwain and M.~Wr{\'o}blewski.
	\newblock Application of quantum k-nn and grover's algorithms for
	recommendation big-data system.
	\newblock In L.~Borzemski, J.~{\'{S}}wi{\k{a}}tek, and Z.~Wilimowska, editors,
	{\em Information Systems Architecture and Technology: Proceedings of 39th
		International Conference on Information Systems Architecture and Technology
		-- ISAT 2018}, pages 235--244, Cham, 2019. Springer International Publishing.
	
	\bibitem{Schuld:2014}
	M.~Schuld, I.~Sinayskiy, and F.~Petruccione.
	\newblock Quantum computing for pattern classification.
	\newblock In {\em PRICAI 2014: Trends in Artificial Intelligence}, 2014.
	
	\bibitem{Shende:2009}
	V.~Shende and I.~L. Markov.
	\newblock On the cnot-cost of toffoli gates.
	\newblock {\em Quantum Information \& Computation}, 9(5):461--486, 2009.
	\newblock ISSN: 1533-7146.
	
	\bibitem{Shor:1999}
	P.~Shor.
	\newblock Polynomial-time algorithms for prime factorization and discrete
	logarithms on a quantum computer.
	\newblock {\em SIAM Review}, 41(2):303--332, 1999.
	\newblock DOI: 10.1137/S0036144598347011.
	
	\bibitem{Steane:1998}
	A.~Steane.
	\newblock Quantum computing.
	\newblock {\em Reports on Progress in Physics}, 61(2):117--173, 1998.
	\newblock DOI: 10.1088/0034-4885/61/2/002.
	
	\bibitem{Stefanowski:2017}
	J.~Stefanowski, K.~Krawiec, and R.~Wrembel.
	\newblock Exploring complex and big data.
	\newblock {\em International Journal of Applied Mathematics and Computer
		Science}, 27(4):669--679, 2017.
	\newblock DOI: 10.1515/amcs-2017-0046.
	
	\bibitem{Trugenberger:2002}
	C.~A. Trugenberger.
	\newblock Quantum pattern recognition.
	\newblock {\em Quantum Information Processing}, 1(6):471--493, 2002.
	\newblock DOI: 10.1023/A:1024022632303.
	
	\bibitem{Veloso:2015}
	B.~Veloso, B.~Malheiro, and J.~C. Burguillo.
	\newblock A multi-agent brokerage platform for media content recommendation.
	\newblock {\em International Journal of Applied Mathematics and Computer
		Science}, 25(3):513--527, 2015.
	\newblock DOI: 10.1515/amcs-2015-0038.
	
	\bibitem{Walther:2005}
	P.~Walther, K.~J. Resch, T.~Rudolph, E.~Schenck, H.~Weinfurter, V.~Vedral,
	M.~Aspelmeyer, and A.~Zeilinger.
	\newblock Experimental one-way quantum computing.
	\newblock {\em Nature}, 434(0):169--176, 2005.
	\newblock DOI: 10.1038/nature03347.
	
	\bibitem{Wiebe:2015}
	N.~Wiebe, A.~Kapoor, and M.~Svore.
	\newblock Quantum algorithms for nearest-neighbor methods for supervised and
	unsupervised learning.
	\newblock {\em Quantum Information and Computation}, 15(3--4):316--356, 2015.
	\newblock URL: = http://dl.acm.org/citation.cfm?id=2871393.2871400.
	
	\bibitem{Wisniewska:2018}
	J~Wi{\'{s}}niewska and M.~Sawerwain.
	\newblock Recognizing the pattern of binary hermitian matrices by quantum knn
	and svm methods.
	\newblock {\em Vietnam Journal of Computer Science}, 5(3):197--204, 2018.
	\newblock DOI: 10.1007/s40595-018-0115-y.
	
\end{thebibliography}

\end{document}